\documentstyle{article}

\begin{document}
\hfill{q-alg/9610004}

\hfill{October, 1996}

\vspace{10mm}

\centerline{\bf   CONTRACTIONS OF INTEGRABLE EQUATIONS  }
\vspace*{0.37truein}
\centerline{\footnotesize N.A.GROMOV, I.V.KOSTYAKOV and V.V.KURATOV}
\vspace*{0.015truein}
\centerline{\footnotesize\it Department of Mathematics, Komi Science Centre}
\baselineskip=10pt
\centerline{\footnotesize\it Ural Division, Russian Academy of Sciences}
\baselineskip=10pt
\centerline{\footnotesize\it Syktyvkar, 167000, Russia}
\baselineskip=10pt
\centerline{\footnotesize\it e-mail: parma@omkomi.intec.ru}
\vspace*{0.225truein}
\vspace*{0.21truein}
\abstract{
The contraction is applied to obtaining of
integrable systems associated with nonsemisimple
algebras. The effect of contraction is splitting
off some components from initial system without loss of
integrability.}
\section{Introduction}
\noindent
The Lie algebraic approach to integrable systems is well known.
There are many models associated with simple finite and infinite
algebras. The step toward general position algebras was made
by Lesnov and Saveliev.\cite{1} They used nonsemisimple algebras and the
contraction procedure. Here we investigate
an action of the contraction procedure on
integrable equations. All examples are related to
Toda and KdV systems. We consider the contractions of finite
and affine algebras and $Z_2$-graded contractions of the
Virasoro algebra.

\section{Contraction}

\subsection{Finite algebra}
\noindent
Let $ {\cal G}_1=(V,[.,.]_1) $
and $ {\cal G}_2=(V,[.,.]_2) $ be two Lie algebras constructed
on the vector space
$ V $. $ {\cal G}_2 $ is a contraction of
$ {\cal G}_1 $ if there exists a family
$ {\Phi}_{\varepsilon}$, ${\varepsilon} \in (0,1], $
of invertible linear transformations of $ V $ so that
$
\lim_{{\varepsilon} \rightarrow 0}
\Phi_{\varepsilon}^{-1}[\Phi_{\varepsilon}{x},\Phi_{\varepsilon}{y}]_1=
[x,y]_2, \quad {\forall x,y}  \subset V
$
One can also say that $ {\cal G}_1 $ is a deformation of $ {\cal G}_2 $.
Let $ {\cal G}_1 $, be a simple Lie algebra of rank $r$,
equipped with an invariant scalar product denoted $(  ,  )$. We choose a
Cartan subalgebra with an orthonormal basis $ H_{i} $ and a set of
simple roots $R$
\begin{equation}
[H_{i},H_{i}]_1=0,\quad
[H_{i},E_{\pm\alpha}]_1={\pm\alpha_{i}}E_{\pm\alpha},\quad
[E_{\alpha},E_{-\alpha}]_1=H_{\alpha},
\end{equation}
where $H_{\alpha}=(E_{\alpha},E_{-\alpha})\sum_{i}\alpha_{i}H_{i}$.
Let $R_c$ denote a subset of simple roots.
If we choose ${\Phi}_{\varepsilon}:
E_{\alpha}\rightarrow {\varepsilon}E_{\alpha}$ for $\alpha\in R_c$,
then for ${\cal G}_2$
the last commutators in (1) are
\begin{equation}
[E_{\alpha},E_{-\alpha}]_2=H_{\alpha} \quad  \alpha \notin R_{c}, \quad
[E_{\alpha},E_{-\alpha}]_2=0 \quad  \alpha \in R_{c}.
\end{equation}

\subsection{Infinite algebra}
\noindent
Let $h_{i},e_{i},f_{i}, i=0,1,...r$ are generators of the affine Kac-Moody
algebra $\hat{\cal G}_1$ with Cartan matrix $\hat k_{ij}=\alpha_{j}(h_{i})$
and commutators
$$
[h_i,h_j]_1=0,
$$
$$
[h_i,e_j]_1=\alpha_{j}(h_i)e_j,\quad [h_i,f_j]_1=-\alpha_{j}(h_i)f_j,
$$
$$
ad^{1-\hat k_{ij}}e_{i}e_{j}=0,\quad ad^{1-\hat k_{ij}}f_{i}f_{j}=0,
$$
\begin{equation}
 [e_i,f_j]_1=\delta_{ij}h_i.
\end{equation}
If we choose ${\Phi}_{\varepsilon}:
e_{p}\rightarrow {\varepsilon}e_{p}$ for some $p$,
then for $\hat {\cal G}_2$
the last commutator in (3) is equal zero for $i=p$.

\subsection{Virasoro algebra}
\noindent
Let us split the generators of the Virasoro algebra $Vir_c$
\begin{equation}
[L_n,L_m]=(n-m)L_{n+m}+{c \over 12}n(n^2-1)\delta_{n+m,0}
\end{equation}
into two sets, even
$L_0=2A_0-{c \over 8}, L_{2n}=2A_p$
and odd
$L_{2n+1}=2B_p.$
In $A_p,B_p$ language (4) looked as
$$
[A_p,A_q]=(p-q)A_{p+q}+{2c \over 12}p(p^2-1)\delta_{p+q,0},
$$
$$
[A_p,B_q]=(p-q-{1 \over 2})B_{p+q},
$$
\begin{equation}
[B_p,B_q]=(p-q)A_{p+q+1}+\\
{2c \over 12}(p-{1 \over 2})(p+{1 \over 2})(p+{3 \over 2})\delta_{p+q+1,0}.
\end{equation}
 There is a $Vir_{2c}$ subalgebra in $Vir_c$:
$Vir_c=Vir_{2c}+something$

This is $Z_2$ graded $Vir_c$.
Three $Z_2$-grading contractions are given by the
 matrix $\varepsilon_1$, $\varepsilon_2$, $\varepsilon_3$.\cite{2}
We will use only $\varepsilon_1$:
$\varepsilon_{11}=\varepsilon_{12}=\varepsilon_{21}=1,\varepsilon_{22}=0$.
In this case the last commutator in (5) is equal zero.

\section{Toda Chain}
\subsection{1D Toda chain}
\noindent
The Toda chain\cite{3} is the system with
$r$ degrees of freedom, phase space coordinates
$ (q_i,p_i) \  i=1,\ldots,r, $
Poisson bracket $ \{p_i,q_j \}={\delta}_{ij} $,
and the Hamiltonian
\begin{equation}
h={\displaystyle{1 \over 2}} \sum_{i}{p_i^2}+\sum_{R}{exp(\alpha \cdot q)},
\end{equation}
where $(\alpha \cdot q)=\sum_{i}\alpha_{i}q_{i}$. The equations of motion
\begin{equation}
\left \{  \begin{array}{l}
        \dot q_i=p_i \\
        \dot p_i=-\sum_{i}{\alpha_i} exp(\alpha \cdot q)
        \end{array} \right .
\end{equation}
admit a Lax pair representation
$$
{\dot L}=[L,M]_{1},
$$
$$
L={\displaystyle{1 \over 2}} \sum_{i}{p_i}{H_i}+\\
\sum_{R}n_{\alpha}{e^{{1 \over 2}(\alpha \cdot q)}}(E_{\alpha}+E_{-\alpha}),
$$
\begin{equation}
M=\\
\sum_{R}n_{\alpha}{e^{{1 \over 2}(\alpha \cdot q)}}(E_{\alpha}-E_{-\alpha}),
\quad n^2_{\alpha}=\displaystyle{1 \over {4(E_{\alpha},E_{-\alpha})}},
\end{equation}
where $H_{i},E_{\pm\alpha}$ and $[ , ]_{1}$ are generators and Lie bracket
of ${\cal G}_1$.
Let ${\cal G}_2$ be a contraction of ${\cal G}_1$
defined in previous section.
We call the system (8) in algebra ${\cal G}_2$ the contracted Toda chain.
The equations of motion, Hamiltonian and Lax pair representation are
$$
\left\{  \begin{array}{l}
        \dot p_i=-\sum_{R/R_c} \alpha_i \exp(\alpha \cdot q) \\
        \dot q_i=p_i,
        \end{array} \right.
$$
$$
 h=\sum_i{\displaystyle{1 \over 2}}p_i^2+\sum_{R/R_c} \,exp(\alpha \cdot q),
$$
\begin{equation}
{\dot L}=[L,M]_{2}.
\end{equation}
The $sl(2)$ Toda contracted chain is a free particle. The $sl(3)$ case
leads
to the elimination of some interacting terms in hamiltonian.

\subsection{2D Toda chain}
\noindent
The equation of motion is zero curvature condition\cite{1}
$$
\left [ {\partial \over \partial{z_+}}+\\
A_{+},{\partial \over \partial{z_-}}+A_{-} \right ]_{1}=0,
$$
\begin{equation}
 A_{\pm}=(hu_{\pm})+(E_{\pm}f_{\pm})=\\
\sum_{i=1}^r(h_iu_i^{\pm}+E_{{\pm}i}f_i^{\pm}).
\end{equation}
For $\rho_i=ln{f_i^{+}f_i^{-}}$ we have in ${\cal G}_1$ algebra
\begin{equation}
{\partial^2 \rho_j \over \partial{z_+}\partial{z_-}}=\\
\sum_{i=1}^r{k_{ji}e^{\rho_i}},
\end {equation}
where $k_{ij}$ is $r\times r$ Cartan matrix of ${\cal G}_1$.
Let ${\cal G}_2$ be a contraction of ${\cal G}_1$.
We call the system (11) in ${\cal G}_2$ algebra the contracted 2D
Toda chain. The equations of motion are
\begin{equation}
{\partial^2\rho_j \over \partial{z_+}\partial{z_-}}=\\
\sum_{i=1}^r{k_{ji}^{'}e^{\rho_i}},
\end{equation}
where several columns of Cartan matrix corresponding to contracted roots
$ \alpha \in R_{c}$ are vanished.
Let $R_c=\alpha_{i}$, then the equations are reduced to
$$
 {\partial^2\rho_i \over \partial{z_+}\partial{z_-}}=\\
 F(\rho_1,...\rho_{i-1},\rho_{i+1},...\rho_r),
$$
\begin{equation}
         {\partial^2\rho_l \over \partial{z_+}\partial{z_-}}=\\
         \sum_{j=1}^r{k_{lj}^{'}e^{\rho_j}}, \quad l,j\neq i.
\end{equation}

The contracted component $\rho_i$ has no self-action and split off
from the other components, because there is no $\rho_i$ in r.h.s. of (13).
Therefore,
we may consider the contraction of 2D Toda chain as the transition
between chain with Cartan matrix $k$ and chain with matrix $\tilde{k}$,
where   $(r-1)\times(r-1)$ matrix $\tilde{k}$ is obtained by crossing out of
$i-$th column and $i-$th row from $k$.
 The equations
of contracted system looked as
$$
 {\partial^2\rho_{contr} \over \partial{z_+}\partial{z_-}}=\\
 F(\rho_1,...\rho_{r-1}),
$$
\begin{equation}
 {\partial^2 \rho_l \over \partial{z_+}\partial{z_-}}=\\
 \sum_{j=1}^{r-1}{\tilde{k}_{lj}}e^{\rho_j},
\end{equation}
where $\rho_{contr}$ is contracted component and $l,j=1,2...,r-1.$
All these transitions may be
classified in terms of Dynkin diagrams using the following property:
if the lines connecting any two points are severed, the resulting
diagram is a Dynkin diagram. Thus, $\tilde{k}$ remains a Cartan matrix.
For example,
there are three contractions of $D_4$ chain:
$$
D_4 \rightarrow A_3 +\rho_{contr},
$$
$$
D_4 \rightarrow D_3+\rho_{contr},
$$
\begin{equation}
D_4 \rightarrow A_1+A_1+A_1+\rho_{contr}.
\end{equation}

\subsection{2D affine Toda system}
\noindent
The base formulas for the affine Toda chain are given in.\cite{4}
The equations
of motion is zero curvature condition
$$
[L,\bar{L}]_1=0, \quad L=e^{-\psi}\partial_{x}e^{\psi} +\Lambda, \quad
\bar{L}=\partial_{t}+e^{-\psi}\bar{\Lambda}e^{\psi},
$$
$$
{\partial^2 \psi \over \partial{x}\partial{t}}=\\
\sum_{i=0}^r{h_{i}e^{\alpha_i(\psi)}},
$$
\begin{equation}
\psi=\sum_{i=0}^r{\psi_{i}h_{i}},\quad \Lambda=\\
\sum_{i=0}^r{e_{i}},\quad \bar{\Lambda}=\sum_{i=0}^r{f_{i}},
\end{equation}
where $h_{i},e_{i},f_{i}, i=0,1,...r$ are generators of the affine Kac-Moody
algebra $\hat{\cal G}_1$ with Cartan matrix $\hat k_{ij}=\alpha_{j}(h_{i})$.
We call the system (16) in algebra $\hat {\cal G}_2$ the contracted  affine
Toda chain.
The Lax par representation and equations of motion are
$$
[L,\bar L]_2=0,
$$
\begin{equation}
{\partial^2\psi_i \over \partial{x}\partial{t}}=\\
\exp{\sum_{i=0}^r{\hat k_{ji}{\psi_j}}} \quad i\neq p ,
\quad {\partial^2\psi_p \over \partial{x}\partial{t}}=0,
\end{equation}
where $p$-th column of Cartan matrix
corresponding to contracted field $\psi_p$
is not arise in equations. Let $\phi_j=\psi_j-\psi_p$
for $j\neq p$ then for $\hat A_n$ (17)  looked as
\begin{equation}
{\partial^2\phi_i \over \partial{x}\partial{t}}=\\
\exp{\sum_{i=0}^{r-1}{\tilde{k}}_{ji}\phi_j},
\quad {\partial^2\psi_{contr} \over \partial{x}\partial{t}}=0,
\end{equation}
where matrix $\tilde{k}$ is obtained by crossing out of
$p-$th column and $p-$th row from $\hat k$. The affine Cartan matrix of
$\hat A_n$ algebra reduced after these crossing to Cartan matrix of
finite algebra. Thus, the affine Toda chain reduced to usual Toda chain.
For example, there are two contractions of $\hat A_3$ chain:
$$
\hat A_3 \rightarrow A_3+\psi_{contr},
$$
\begin{equation}
\hat A_3 \rightarrow D_3+\psi_{contr}.
\end{equation}
There is  a contraction of
$\hat A_2^{2}$ chain, $\hat A_2^{2}\rightarrow A_1+\psi_{contr}$, which
described a transition from the Sine-Gordon to the Liouville equation.

Thus, contractions give a method of transitions between the
different Toda models.

\section{KdV equation}

\subsection{The AKNS hierarchy}
\noindent
 One of the approach to the contraction of the KdV equation is based
 on AKNS\cite{5} method of reproducing of integrable equations.
 It is easily shown for $sl(2)$ AKNS hierarchy that after contraction
 to $isl(2)$ algebra all equations are reduced to linear equations.

\subsection{The second Poisson bracket structure}
\noindent
 The another approach to the contraction of the KdV equation
 is based on $Z_2$-graded contraction of the Virasoro algebra.
 The KdV equation can be written as\cite{5}
\begin{equation}
 u_t=(\partial^3+2u\partial+2\partial u){\delta H \over \delta u} =\\
 \Theta{\delta H \over \delta u},
\end{equation}
where $ \displaystyle{H=I_{2}[u]=\int {u^2 \over 2}dx,
{\delta \over \delta u}}$ denotes a variational derivative.
The associated Poisson bracket for functionals $F[u]$ and $G[u]$ are
\begin{equation}
 \{F[u],G[u]\}=\\
 \int {\delta F \over \delta u} \Theta {\delta G \over \delta u} dx.
\end{equation}
The
conserved quantities $I_n[u]$ of KdV equation are in involution
with respect to this bracket.
For $u(x)$ coordinats Poisson bracket looked as
\begin{equation}
 \{u(x),u(y)\}=\Theta \delta (x-y).
\end{equation}
Considering equating (22) in the Fourier space, one is lead
to the Virasoro algebra
\begin{equation}
\{L_n,L_m\}=(n-m)L_{n+m}+{c \over 12}n(n^2-1)\delta_{n+m,0}.
\end{equation}
After rewriting in $Z_2$ grading language (sect.2.3), and returning from the
Fourier to configuration space we have two kinds
of coordinates $v(x)$, $w(x)$
and a new $2 \times 2$ matrix $\hat \Theta-$operator.
It is easily shown that after the $\varepsilon_1$ contraction
of the Virasoro algebra,
$\hat \Theta$-operator looked as
\begin{equation}
 \hat \Theta_{contr}= \left (\begin{array}{cc}
        \partial^3+2v\partial+2\partial v  & 2w\partial+2\partial w+iw/2 \cr
        2w\partial+2\partial w-iw/2        &      0
            \end{array} \right ).
\end{equation}
We can define the contracted Poisson bracket
\begin{equation}
\{F[v,w],G[v,w]\}_{contr}=\\
\int\nabla F[v,w]\hat \Theta_{contr} \nabla G[v,w]dx,
\end{equation}
where $\nabla$ denotes variational gradient.
This bracket has at least one set
of the conserved quantities $I_n[v]$ in involution,
because there is a Virasoro subalgebra
in contracted Virasoro algebra. Put $H[v,w]=I_{2}[v]$ as the
Hamiltonian than we have two equations
$$
v_t=\{v,H\}_{contr}=\partial^3v+6vv_x,
$$
\begin{equation}
w_t=\{w,H\}_{contr}=2vw_x+4v_xw-{i \over 2}vw.
\end{equation}
Thus, after contraction we have again the KdV equation for $v(x)$
and some equation for $w(x)$.

\end{document}